\documentclass{article}
\usepackage{spconf,amsmath,graphicx}
\usepackage{cite}
\usepackage{amssymb,amsfonts}
\usepackage{enumitem}
\usepackage{bm}
\usepackage{graphicx}
\usepackage{textcomp}
\usepackage{xcolor}
\usepackage{booktabs}
\usepackage[hidelinks]{hyperref}
\usepackage{multirow}
\usepackage{tabularx}

\newcommand{\X}{\boldsymbol{X}}

\newcommand{\Sb}{\boldsymbol{S}}
\usepackage{algorithmic}
\usepackage[ruled,vlined]{algorithm2e}

\title{Supervised Hierarchical Clustering using Graph Neural Networks for Speaker Diarization}
\name{Prachi Singh \thanks{This work was supported by the grants from  the British Telecom Research Center.}, Amrit Kaul, Sriram Ganapathy}
\address{LEAP Lab, Electrical Engineering,
Indian Institute of Science,Bangalore.\\
\texttt{\small{prachisingh@iisc.ac.in}}}

\begin{document}
%
%
%

\ninept
\maketitle
\begin{abstract}
Conventional methods for speaker diarization involve windowing an audio file into short segments to extract speaker embeddings, followed by an unsupervised clustering of the embeddings. 
This multi-step approach generates speaker assignments for each segment. 
In this paper, we propose a novel Supervised HierArchical gRaph Clustering algorithm (SHARC) for speaker diarization where we introduce a hierarchical structure using Graph Neural Network (GNN) to perform supervised clustering. The supervision allows the model to update the representations and directly improve the clustering performance, thus enabling a single-step approach for diarization. In the proposed work, the input segment embeddings are treated as nodes of a graph with the edge weights corresponding to the similarity scores between the nodes.
We also propose an approach to jointly update the embedding extractor and the GNN model to perform end-to-end speaker diarization (E2E-SHARC). 
During inference, the hierarchical clustering is performed using node densities and edge existence probabilities to merge the segments until convergence. 
In the diarization experiments, we illustrate that the proposed E2E-SHARC approach achieves $53\%$ and $44\%$ relative improvements over the baseline systems on benchmark datasets  like AMI and Voxconverse, respectively.
\end{abstract}
\section{Introduction}
\label{sec:intro}
Speaker Diarization (SD) is the task of segmenting an audio file based on speaker identity. The task has important applications in rich speech transcription for multi-speaker conversational audio like customer call center data, doctor patient conversations and meeting data.

The conventional approach for the task of SD involves multiple steps. In the first step, the audio is windowed into short segments (1-2 s) and fed to a speaker embedding extractor. The speaker embedding extractors are deep neural networks trained for the speaker classification task.  The output of penultimate layer, called as embeddings, provides a good initial speaker representation (for example, x-vectors) \cite{snyder2019speaker,snyder2018x}. In a subsequent step, these speaker embeddings are clustered based on similarity scores computed using methods like  Probabilitic Linear Discriminant Analysis (PLDA) scoring \cite{ioffe2006probabilistic,pldaivec}. 
The most common clustering approach is the agglomerative hierarchical clustering (AHC) \cite{AHC}, which merges two clusters at each time step based on similarity scores until the required number of clusters/speakers are attained. Other approaches involve spectral clustering (SC) \cite{wang2018speaker},  k-means clustering \cite{likas2003global} and graph based clustering \cite{PIC,SSPIC}. 

Recently, the end-to-end neural diarization~\cite{shinji2019ASRU, shinji2020Interspeech} approaches involving transformers have proved effective in handling overlaps. However, due to the difficulty in handling more than $4$ speakers, pairwise metric learning loss is proposed recently \cite{horiguchi2021towards}. There have been recent efforts on clustering algorithms
to improve the diarization performance over the conventional approach. A graph-based agglomerative clustering called path integral clustering proposed by Zhang et al. \cite{PIC} is shown to outperform other clustering approaches on CALLHOME 
 and AMI datasets \cite{SSPIC}. Similarly, metric learning approaches are introduced in \cite{wang2018speaker,Lin2019} to improve the speaker similarity scores. 
In a recent work, Singh et al. \cite{SSPIC, singh2021self} introduced self-supervised metric learning using clustering output labels as the pseudo labels for model training. 

Most of the previous approaches for diarization are trained to improve the similarity scores. However, they still use an unsupervised clustering algorithm to obtain the final labels. We hypothesize that this limits their performance as they are not trained with the clustering objectives. On the other hand, EEND models require a large amount of data and hundreds of hours of training. We propose a simple approach to SD which is not data intensive and can handle large number of speakers (more than 7) while training and evaluation. The approach is called as Supervised HierArchical gRaph Clustering algorithm (SHARC). Our work is inspired by Xing et al. \cite{HGNN}, where a supervised learning approach to image clustering was proposed.  
We perform supervised representation learning and clustering jointly without requiring an external clustering algorithm. The major contributions are: \vspace{-4pt}
\begin{enumerate}
\item Introducing supervised hierarchical clustering using Graph Neural Networks (GNN) for diarization.\vspace{-4pt}
\item Developing the framework for joint representation learning and clustering using supervision.\vspace{-4pt}
\item \textcolor{black}{Achieving state-of-the-art performance on two benchmark datasets.} 
\end{enumerate}
\begin{figure*}
\centering
\includegraphics[width=14cm, trim={0cm 0cm 0cm 0cm},clip]{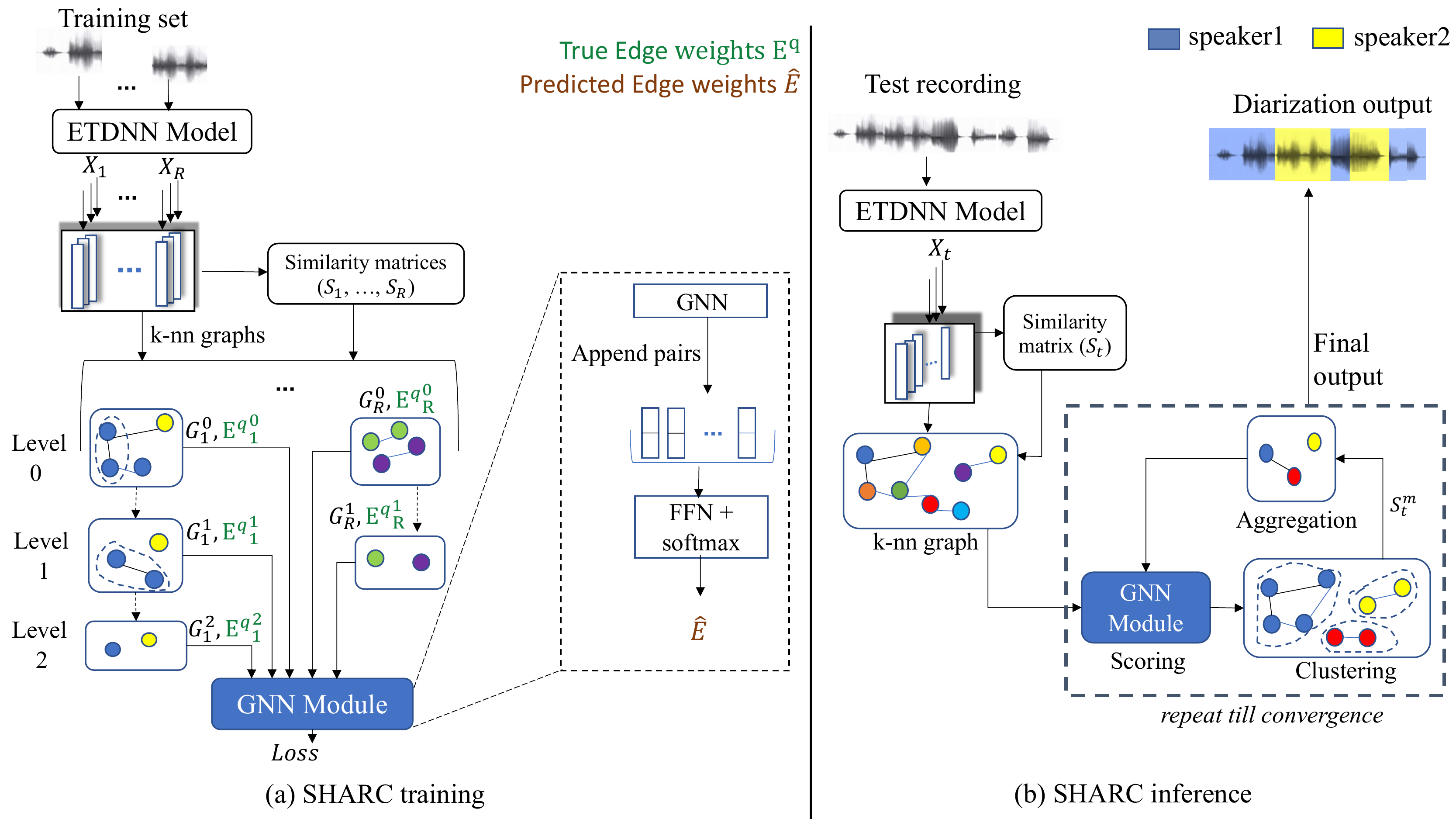} 
\caption{Block diagram of proposed SHARC method. The ETDNN model and GNN are the extended time delay network model for x-vector extraction and the graph neural network for score prediction. FFN stands for feed forward network. The left side (a) shows the training steps and the right side (b) shows the inference steps. }
\label{fig:blockdiagram}
\end{figure*}
\vspace{-6pt}
\section{Related Work and Background}
\label{sec:relwork}
This section highlights previous works on SD by representing multi-speaker audio file in the form of a graph. 
 We first introduce GNN and their use in metric learning and supervised clustering in other domains. Then, we describe a variant of GNN, called GraphSAGE \cite{hamilton2017inductive} which is used in our approach.

Wang et al. \cite{GNNfordiarization} proposed GNN for metric learning. The inputs to the model are x-vectors/d-vectors and the PLDA similarity score. The output of the model is a probability score of whether two nodes are connected or not. 
The Graph Convolution Network (GCN) \cite{kipf2016semi}, the most common variant of GNNs, are used in \cite{TongGCN} for semi-supervised training using clustering output as ``pseudo-labels". \newline 

\textbf{GraphSAGE:} The GCN is inherently transductive and does not generalize to unseen nodes. The GraphSAGE\cite{hamilton2017inductive}, another variant of GNN, is a representation learning technique suitable for dynamic graphs. It can predict the embedding of a new node without requiring a re-training procedure. The GraphSAGE learns aggregator functions that can induce the embedding of a new node given its features and neighborhood. 
First, a graph is constructed using the embeddings as the nodes. The edges are connected using the similarity scores between the embeddings.
Instead of training individual embeddings for each node, a function is learnt that generates embeddings by sampling and aggregating features from a node’s local neighborhood. The aggregate function outputs a single neighborhood embedding by taking a weighted average of each neighbor’s embedding. 
\vspace{-4pt}
\section{Proposed Approach}
\label{sec:proposedapproach}
The  Supervised HierArchical gRaph Clustering algorithm (SHARC) model is shown in Figure \ref{fig:blockdiagram}. It introduces a hierarchical structure in the GNN-based clustering. Figure \ref{fig:blockdiagram}(a), shows the training procedure using $R$ audio recordings $r\in\{1, 2, .., R\}$ where $r$ is the recording-id assigned to each recording in the dataset. It involves extracting short segment embeddings such as x-vectors $\boldsymbol{\mathcal{X}}=\{\X_1, \X_2, ...,\X_R\}$  from an Extended Time Delay Neural Network (ETDNN) \cite{snyder2019speaker}
 for all recordings where $\X_r \in \mathcal{R}^{N_r\times F}$, 
 $N_r$ is the number of x-vectors for recording $r$ and $F$ is the dimension of x-vectors. These are used to form graphs at different levels of hierarchy denoted as $G= \{G^0_{1}, G^0_{2}, ..., G^1_{1}, ..., G^{Mr}_{R}\}$ where $G^m_r$ is a graph of recording $r$ at level $m$ and $M_r$ is the maximum number of levels created for $r$. The nodes of the graphs are obtained from $\boldsymbol{\mathcal{X}}$, and edges are connected using $k$-nearest neighbors with weights coming from similarity matrices $\boldsymbol{\mathcal{S}^m}=\{\Sb^m_1, ...,\Sb^m_R\}$ for level $m$ where, $\Sb^m_r \in \mathcal{R}^{N^m_r\times N^m_r}$, $N^m_r$ is number of nodes at level $m$ for recording $r$. The graphs are constructed at different clustering levels by merging the node features of each cluster and recomputing the similarity matrix, as discussed in Section \ref{sec:graphgeneration}.
For training, a set of graphs $G$ are fed to the GNN module in batches. The module comprises of GNN along with a feed forward network to predict edge  weights  $\hat{E_m} \in \mathcal{R}^{N^m_r\times k}$ of all nodes with their k-nearest neighbors. The loss is computed using $E^q$  (true edge weights) and $\hat{E_m}$ (predicted edge weights). The details of the GNN scoring and loss computation are given in Section  \ref{sec:training}.

Figure \ref{fig:blockdiagram}(b), shows the inference block diagram. For a test recording $t$, x-vectors $\X_t$ and $\Sb_t$ are extracted and a graph $G^0_t$ is constructed at level 0. Then it is passed to the clustering module which iteratively performs clustering using edge predictions from GNN module followed by merging nodes of same cluster and then, reconstructing graph for next level $m$. 
This process stops if the graph has no edges ($G^m=\{\phi\}$) or maximum allowed level $M$ is attained.
The algorithm outputs cluster labels predicted for the nodes at the top level, propagating down to the original embeddings. The process is summarized in Algorithm \ref{algo:sharc}. 
\vspace{-8pt}
\subsection{Graph generation}
\label{sec:graphgeneration}
In this step, a hierarchy of graphs, $G^m_r=(V^m,E_m)$, is created using $\X_r$ and $\Sb^m_r$ where $V^m=\{v_1, v_2, ..., v_{n}\}$ is the set of the nodes and $E_m$ is the set of the edges. Each graph consists of  node representations $H^m_r = \{h_1^{(m)}, h_2^{(m)},...,h_n^{(m)}\} \in \mathcal{R}^{F'Xn}$ where $n=N^m_r$ is the number of nodes at level $m$.
$E_m$ is obtained using $\Sb^m_r \in [0,1]$ considering $k$-nearest neighbors of each node in $V^m$.
At level $m=0$, we consider each embedding as individual cluster. Therefore,  node representations are given as $H^0_r=[\X_r;\X_r]$.  
For any level $m>0$, the  node representation is obtained by concatenating the identity feature and the average feature of the current cluster, as described in Section \ref{sec:feat_aggre}. 
\vspace{-6pt}
\subsection{\textcolor{black}{GNN scoring and clustering}}
\label{sec:clustering}
\textcolor{black}
{The node representations $H^m$ at each level  $m$ are passed to the GNN scoring function $\Phi$. It predicts edge linkage probability ($p_{ij}$) which indicates presence of an edge $(v_i,v_j)\in V^m$ along with node density ($\hat{d_i}$) which measures how densely the node is connected with its neighbors. }
After GNN scoring, the clustering is performed. 
At each level of hierarchy $m$, it creates a candidate 
edge set $\varepsilon (i)$, for the node $v_i$, with edge connection threshold $p_\tau$, as given below.
\begin{equation}
\label{eqn:edge_set}
\varepsilon (i) = \{j|(v_i,v_j) \in E_m, \quad \hat{d}_i \leq \hat{d}_j \quad \textrm{and}\quad p_{ij} \geq p_\tau \} 
\end{equation}
For any $i$, if $\varepsilon(i)$ is not empty, we pick $j = $ argmax$_{j \in \varepsilon (i)} \hat{e}_{ij}$ and add $(v_i,v_j)$ to $E'_m$ where $ \hat{e}_{ij}$ is the predicted edge weights, given as,
 \begin{equation}
 \hat{e}_{ij} = 2p_{ij}-1\in [-1,1]\forall j\in N_i^k
 \end{equation}
 Here, $N_i^k$ are the k-nearest neighbors of node $v_i$.  After a full pass over every node, $E'_m$ forms a set of connected components $C'_m$, which serves as the designated clusters for the next level ($m+1$). 
The clustering stops when there are no connected components present in the graph.
\vspace{-6pt}
\begin{algorithm}[t!]
    \SetAlgoLined
    \textbf{Initialize:}  $M \gets$ maximum no. of levels;
    $m \gets 0$; \\
    $k \gets $ no. of nearest neighbors; 
     $H^0  \gets [\X;\X] $\\
         \While{$m\le M$}{
      \begin{enumerate}
         \item Graph generation: $G^m \gets graph (H^{m},  \Sb, k)$ \\
  
             \textbf{If} ($G^m \gets \{\phi\}$) : $M=m$; $break$
	 \vspace{-2pt}
          \item GNN Scoring: $\hat{E}_m \gets \Phi(G^m,H^{m})$\\
           \vspace{-2pt}
          \item $C_m' \gets Clustering(\hat{E}_m)$
           \vspace{-2pt}
          \item Aggregation: $H^{m+1} \gets \Psi(H^{m},C_m')$\\
           \vspace{-2pt}
          \item $m \gets m+1$
          \end{enumerate}
     }
     \textbf{Output:}\vspace{0.5em}  Predicted $\boldsymbol{\hat{Y}}=\{\hat{y}_1,..,\hat{y}_N\}$ using $C'_{\{1:m-1\}}$ 
     \caption{\textcolor{black}{SHARC Inference}}
    \label{algo:sharc}
    \vspace{-5pt}
    \end{algorithm}

\subsection{Feature aggregation} 
\label{sec:feat_aggre}
To obtain  node representations for next level $H^{m+1}$, the  connected components $C'_m$ obtained from the clustering along with $H^m$ are passed to an  aggregation function $\Psi$  . The function $\Psi$ concatenates identity feature ${\Tilde{h}_i}^{(m+1)}$ and average feature ${\Bar{h}_i}^{(m+1)}$ of each cluster $i$ to obtain ${h_i}^{(m+1)} = [{\Tilde{h}_i}^{(m+1)};{\Bar{h}_i}^{(m+1)}]$.   The identity feature of node $i$ at level $m+1$ is the  feature of the node  which has highest node density at level $m$ in the cluster $i$. The average feature is computed by taking average of all the identity features from previous level of that cluster, given as,
\begin{equation}
\label{eqn:clusterfeat}
\begin{split}
{\Tilde{h}_i}^{(m+1)} = {\Tilde{h}^{(m)}}_{z_i} ; \quad \quad
{\Bar{h}_i}^{(m+1)} = \frac{1}{|{c_i}^{(m)}|}\sum_{j \in {c_i}^{(m)}} {\Tilde{h}_j}^{(m)}
\end{split}
\end{equation}
where $z_i$ = argmax$_{j \in {c_i}^{(m)}} {\hat{d}_j}^{(m)}$.
\vspace{-4pt}
\subsection {\textcolor{black}{GNN module architecture and training}}
\label{sec:training}
\textcolor{black}{GNN scoring function $\Phi$ is implemented as a  learnable GNN module designed for supervised clustering. 
The module consists of one GraphSAGE\cite{hamilton2017inductive} layer with $F'=2048$ neurons.  
Each graph $G^m_r$, 
 containing source and destination node pairs, is fed to the GNN module. 
It takes  node representations $H^{m}$
 and their edge connections as input and generates latent representations denoted as $\hat{H}^{(m)} \in \mathcal{R}^{F'\times n}$,
with $n$ being the number of embeddings at a level m. The pair of embeddings are concatenated $[\hat{h}_i;\hat{h}_j]$ and passed to a three-layer fully connected feed-forward network with a size of $\{1024,1024,2\}$ followed by softmax activation to generate linkage probability $p_{ij}$.  The predicted node density is  computed as:
\vspace{-5pt}
\begin{equation}
    \hat{d}_i=\frac{1}{k}\sum_{j \in N_i^k}\hat{e}_{ij}\Sb_r(i,j)
\end{equation}
The ground truth density $d_i$ is obtained using ground truth edge weight $e^q_{ij}= 2q_{ij} - 1\in E^q \{-1,1\}^{N_rXk}$,  
where $q_{ij}=1$ if nodes $v_i$ and $v_j$ belong to the same cluster, otherwise $q_{ij}=0$.  A node with higher density is a better representative of the cluster than a node with lower density. }
{\textcolor{black}{Each node $v_i$ has a cluster (speaker) label $y_i$ in the training set, allowing the function to learn the clustering criterion from the data. The loss function for training is given as follows:
\vspace{-6pt}
\begin{equation}
\label{eqn:tot_loss}
L = L_{conn} + L_{den}
\end{equation}
where $L_{conn}$ is the pairwise binary cross entropy loss based on linkage probability across all the possible edges in $E$ accumulated across all levels and recordings in a batch. 
$L_{den}$ represents mean squared error  (MSE)  loss between ground truth node density $d_i$ and predicted node density $\hat{d}_i$ $\forall i \in \{1, ..., |V|\}$, where $|V|$ is the cardinality of $V$.}

\vspace{-5pt}
\subsection{E2E-SHARC}
The SHARC model described in the previous section also allows the computation of gradients of the loss function w.r.t the input x-vector embeddings. The computation of these gradients enables the fine-tuning of the embedding extractor.
We remove the classification layer from the 13-layer ETDNN model \cite{zeinali2019but} and connect the $11^{th}$ affine layer output with the SHARC model input. This model is trained using $40$-D mel-spectrogram feature vectors and  similarity matrices as input. 
The details of ETDNN embedding extractor are described in Section \ref{sec:baseline}.
 The training loss is the same as the SHARC model (Equation \ref{eqn:tot_loss}). 
 This approach is referred as End-to-End Supervised HierArchical gRaph Clustering (E2E-SHARC). 
 \vspace{-6pt}
\section{Experiments}
\vspace{-6pt}
\label{sec:exp}

\subsection{Datasets}
\label{sec:datasets}
\subsubsection{\textcolor{black}{AMI}}
\vspace{-2pt}
\begin{itemize}[leftmargin=*]
	\item \textbf{Train, dev and eval sets}: 
	The AMI dataset~\cite{mccowan2005ami} contains meeting recordings from four different sites (Edinburgh, Idiap, TNO, Brno). 
It comprises of training, development (dev) and evaluation (eval) sets consisting of $136$, $18$ and $16$ recordings sampled at $16$kHz, respectively.
	The number of speakers and the duration ranges of each recording from 3-5 and $20$-$60$ mins, respectively.
	\end{itemize}
\vspace{-5pt}
\subsubsection{\textcolor{black}{Voxconverse}}
\vspace{-2pt}
\begin{itemize}[leftmargin=*]

\item \textbf{Train set}: The dataset used for training Voxconverse model is simulated using Voxceleb 1 and 2 \cite{nagrani2017voxceleb,Chung2018} and Librispeech \cite{panayotov2015librispeech} using the recipe from \cite{shinji2019ASRU}.
We simulated $5000$ mixtures containing $2$-$5$ speakers with duration ranging from $150$-$440$ s. This generates $1000$ hrs of data with $6,023$ speakers. 
	\vspace{-2pt}
	
	\item \textbf{Voxconverse dev and eval sets}: It is an audio-visual diarization dataset\cite{chung20_interspeech} consisting of multispeaker human speech recordings extracted from YouTube videos. It is divided into a development (dev) set and an evaluation (eval) set consisting of $216$ and $232$ recordings respectively. The duration of a recording ranges from $22$-$1200$ s. The number of speakers per recording varies from $1$-$21$.
\end{itemize}
\begin{table}[]
\caption{Choice of hyper-parameters for train, dev, eval split of AMI and Voxconverse datasets. The parameters $k^*$ and $p_{\tau}^*$ are used in E2E-SHARC training.}
\label{tab:hyperparameters}
\centering
\resizebox{0.85\columnwidth}{!}{
\begin{tabular}{|l|lll|lll|}
\hline
\multirow{2}{*}{Parameters} & \multicolumn{3}{c|}{AMI}                                     & \multicolumn{3}{c|}{Voxconverse}                              \\ \cline{2-7} 
                            & \multicolumn{1}{l|}{Train} & \multicolumn{1}{l|}{Dev} & Eval & \multicolumn{1}{l|}{Train} & \multicolumn{1}{l|}{Dev} & Eval \\ \hline
  $k$                           & \multicolumn{1}{l|}{60}    & \multicolumn{1}{l|}{60}  & 60   & \multicolumn{1}{l|}{60}    & \multicolumn{1}{l|}{30}  &30   \\
$p_{\tau}$                & \multicolumn{1}{l|}{-}     & \multicolumn{1}{l|}{0.0} & 0.0  & \multicolumn{1}{l|}{-}     & \multicolumn{1}{l|}{0.5} &0.8  \\
$k^*$                     & \multicolumn{1}{l|}{30}    & \multicolumn{1}{l|}{50}  & 50   & \multicolumn{1}{l|}{60}    & \multicolumn{1}{l|}{30}  & 30   \\
$p_{\tau}^*$                  & \multicolumn{1}{l|}{-}     & \multicolumn{1}{l|}{0.0} & 0.0  & \multicolumn{1}{l|}{-}     & \multicolumn{1}{l|}{0.9} & 0.8  \\ \hline
\end{tabular}
}
\vspace{-5pt}
\end{table}
\begin{table}[t!]
		\caption{\color{black}{DER (\%) comparison on the AMI SDM and  Voxconverse datasets with the baseline methods. OVP: overlap, COL: collar.}}
		\vspace{0.2cm} 
		\label{tab:ami_unknown}
		\centering
		\resizebox{\columnwidth}{!}{\begin{tabular}{|l|c|c|c|c|} 
				\hline
				
				\multirow{2}{*}{\textbf{AMI SDM System}} & \multicolumn{2}{c|}{\textbf{with OVP + no COL}}& \multicolumn{2}{c|}{\textbf{w/out OVP + COL}} \\ 
				\cline{2-5}
				& Dev.   & Eval. & Dev. & Eval.  \\ 
				\hline
				
				x-vec + PLDA + AHC \cite{ryant21_interspeech}   &24.50 &29.51 &7.61  &14.59   \\ 
				x-vec + PLDA + SC   &19.8  &22.29 & 4.1 &5.76  \\ 
				x-vec + PLDA + SHARC   &$\textbf{19.71}$  &21.44  & $\textbf{3.91}$ & 4.88   \\ 
				E2E-SHARC   &20.59 &$\textbf{19.83}$ &5.15  &$\textbf{2.89}$   \\ 
				 $\quad \quad -\quad \quad$ + VBx \cite{landini2020but}  &$\textbf{19.35}$  &$\textbf{19.82}$  &$\textbf{3.46}$  &$\textbf{2.73}$   \\ 
				\hline
%
  				 \multicolumn{5}{|l|}{\textbf{Voxconverse System}}
				 \\ 
				\hline
				
%
				x-vec + PLDA + AHC \cite{ryant21_interspeech}   &12.68 &13.41 &7.82  &9.28   \\ 
				x-vec + PLDA + SC   &10.78  &14.02 &6.52  &9.92  \\ 
				x-vec + PLDA + SHARC   &10.25  & 13.29 &6.06  &9.40   \\ 
				E2E-SHARC   &$\boldsymbol{9.90}$ &$\boldsymbol{11.68}$ &$\boldsymbol{5.68}$  &$\boldsymbol{7.65}$  \\ 
				$\quad \quad - \quad \quad$+ VBx \cite{landini2020but}    &$\boldsymbol{8.29}$ &$\boldsymbol{9.67}$ &$\boldsymbol{3.94}$  &$\boldsymbol{5.51}$  \\ 
								\hline
		\end{tabular}}
		\vspace{-0.4cm}
	\end{table}
\vspace{-15pt}
\subsection{Baseline system}
\vspace{-2pt}
\label{sec:baseline}
The baseline method is an x-vector-clustering based approach followed in \cite{ryant21_interspeech,singh2021self}.
First, the recording is divided into 1.5 s short segments with 0.75 s shift. The 40-D mel-spectrogram features are computed from each segment which is passed to the ETDNN model \cite{snyder2019speaker} to extract 
512-D x-vectors. The ETDNN model  follows the Big-DNN architecture described in \cite{zeinali2019but} and is trained on the VoxCeleb1 \cite{nagrani2017voxceleb} and VoxCeleb2 \cite{Chung2018} datasets, for speaker identification task, to discriminate among the $7,146$ speakers. 
The whitening transform, length normalization and recording level PCA are applied to the x-vectors as pre-processing steps 
to compute the PLDA similarity score matrix and  perform clustering to generate speaker labels for each segment. 
For comparison, we have used two most popular clustering approaches - AHC \cite{AHC} and SC \cite{spectral_clustering}. To perform AHC, the PLDA is used directly. For SC, we convert the scores in $[0,1]$ range by applying sigmoid with temperature parameter $\tau=10$ (best value obtained from experimentation).
\vspace{-10pt}
\subsection{\textcolor{black}{Training configuration}}
For training the SHARC model, we extract x-vectors with a window of duration 1.5 s with 0.75 s shift, from single speaker regions of the training set. The similarity score matrices, $\Sb^m$, are obtained using baseline PLDA models which are fed to the GNN module described in Section \ref{sec:training}.
The possible levels of each recording depend on the number of x-vectors ($N_r$) and the choice of $k$. 

To train the end-to-end SHARC model, the weights of the x-vector model are initialized with the pre-trained ETDNN model while the SHARC model weights are initialized with the one obtained from SHARC training. The input to the model is 40-D mel-spectrogram computed from 1.5 s with 0.75 s shift. 
To prevent overfitting of the embedding extractor, the pre-trained x-vectors are added to the embedding extractor output before feeding to the GNN.
\vspace{-15pt}
\subsection{Choice of hyper-parameters}
\label{hyperparameters}
The SHARC model is trained with Stochastic Gradient Descent (SGD) optimizer with a learning rate $lr$=$0.01$ (for Voxconverse) and $lr$=$0.001$(for AMI) for $500$ epochs. Similarly, the E2E-SHARC is also trained with an SGD optimizer. In this case, the learning rate is $1e$-$06$ for the ETDNN model and $1e$-$03$ for the SHARC model, trained for 20 epochs.
The hyperparameters $k, p_{\tau}$ are selected based on the best performance on the dev set for the eval set and vice versa. \textcolor{black}{The maximum number of levels $M$ is initially set to $15$ to avoid infinite loops but the algorithm converges at $M\le3$}.
Table \ref{tab:hyperparameters} shows the values of hyperparameters obtained for the AMI and Voxconverse datasets.

%
%
%
%
%
	
\begin{table}[t]
		\caption{\color{black}{DER (\%, w/out overlap $+$ with collar) comparison with state-of-the-art on AMI MDM (without TNO sets) and Voxconverse datasets. 
}}
		\vspace{0.2cm} 
		\label{tab:amivox}
		\centering
		
		\resizebox{0.9\columnwidth}{!}{
		\begin{tabular}{|l|c|c|} 
				\hline
				
				\multirow{1}{*}{\textbf{AMI MDM System}} 
				& Dev.   & Eval.  \\ 
				\hline
				
				x-vec(ResNet101)+AHC+VBx \cite{landini2020bayesian}  &2.78 &3.09 \\
				ECAPA-TDNN  \cite{dawalatabad2021ecapa} &3.66  &$\boldsymbol{3.01}$   \\ 
				SelfSup-PLDA-PIC (+VBx)\cite{singh2021self}&5.38 ($\boldsymbol{2.18}$) &4.63 (3.27) \\
				SHARC (+VBx)   &3.58 (3.72) &$\textbf{2.29 (2.11)}$    \\ 
												\hline
			       \multirow{1}{*}{\textbf{Voxconverse System}} 
				& Dev.   & Eval.  \\ 
				\hline
				
				Voxconverse challenge \cite{chung20_interspeech}  &24.57 &$-$ \\
				VBx BUT system \cite{landini2021analysis}  &4.36 &$-$ \\
				 Wang et. al. \cite{wang2022similarity}  &4.41 &5.82 \\
E2E-SHARC +VBx   &$\boldsymbol{3.94}$ &$\boldsymbol{5.51}$    \\ 
				\hline
		\end{tabular}}
		\vspace{-10pt}
	\end{table}
\vspace{-15pt}
\section{Results}
\vspace{-10pt}
\label{sec:results}
The proposed approaches are evaluated using the diarization error rate (DER) metric \cite{ryant21_interspeech}. 
In our work, we use ground truth speech regions for performance comparison. The DERs are computed for two cases. The first case considers overlaps and without collar regions, and the second case ignores overlaps and incorporates a tolerance collar of 0.25 s.
Table \ref{tab:ami_unknown} shows that the proposed SHARC model improves over the baseline systems, and the performance further improves with the E2E-SHARC model for both datasets. To incorporate temporal continuity, we applied a re-segmentation approach using Variational Bayes inference (VBx) \cite{landini2020but} with the E2E-SHARC clustering labels as initialization, which further boosted the performance. As shown in Table \ref{tab:ami_unknown}, for the AMI SDM dataset, we obtain $15.6\%$ and $52.6\%$ relative improvements for the dev and eval set, respectively over the PLDA-SC baseline (best). Similarly, we achieve $39.6\%$ and $44.4\%$ relative improvements over the Voxconverse baseline (PLDA- SC) for the dev and eval set, respectively. 

Table \ref{tab:amivox} compares proposed approach performance with state-of-the-art systems. \textcolor{black}{The widely reported beamformed AMI multi-distant microphone (MDM) dataset, without TNO recordings, is used for benchmarking. The beamformed recordings are obtained using} \cite{anguerabeamformit}.
The proposed SHARC model has the lowest DER for eval set compared to all previous SOTA approaches. 
For the Voxconverse dataset, we compare it with the challenge baseline and other published results. 
Here, the E2E-SHARC with VBx shows the best results compared to previously published results.
\vspace{-10pt}
\section{Summary}
\vspace{-6pt}
\label{sec:summary}
We have proposed a supervised hierarchical clustering algorithm using graph neural networks for speaker diarization. The GNN module learns the edge linkages and node densities across all levels of hierarchy. The proposed approach enables the learnt GNN module to perform clustering hierarchically based on merging criteria which can handle a large number of speakers. The method is further extended to perform end-to-end diarization by jointly learning the embedding extractor and the GNN module. With challenging diarization datasets, we have illustrated the performance improvements obtained using the proposed approach.
\vspace{-12pt}
\section{Acknowledgements}
\vspace{-5pt}
The authors would like to thank Michael Free, Rohit Singh, Shakti Srivastava of British Telecom Research for their valuable inputs. 

\bibliographystyle{IEEEbib}
\bibliography{Template_revised}

\begin{thebibliography}{10}

\bibitem{snyder2019speaker}
David Snyder, Daniel Garcia-Romero, Gregory Sell, Alan McCree, Daniel Povey,
  and Sanjeev Khudanpur,
\newblock ``Speaker recognition for multi-speaker conversations using
  x-vectors,''
\newblock in {\em IEEE ICASSP}, 2019, pp. 5796--5800.

\bibitem{snyder2018x}
David Snyder, Daniel Garcia-Romero, Gregory Sell, Daniel Povey, and Sanjeev
  Khudanpur,
\newblock ``X-vectors: Robust {DNN} embeddings for speaker recognition,''
\newblock in {\em IEEE ICASSP}, 2018, pp. 5329--5333.

\bibitem{ioffe2006probabilistic}
Sergey Ioffe,
\newblock ``Probabilistic linear discriminant analysis,''
\newblock in {\em European Conference on Computer Vision}. Springer, 2006, pp.
  531--542.

\bibitem{pldaivec}
Gregory Sell and Daniel Garcia-Romero,
\newblock ``Speaker diarization with {PLDA} i-vector scoring and unsupervised
  calibration,''
\newblock in {\em IEEE Spoken Language Technology Workshop (SLT)}, 2014, pp.
  413--417.

\bibitem{AHC}
William H.~E. Day and Herbert Edelsbrunner,
\newblock ``Efficient algorithms for agglomerative hierarchical clustering
  methods,''
\newblock {\em Journal of Classification}, vol. 1, pp. 7--24, 1984.

\bibitem{wang2018speaker}
Quan Wang, Carlton Downey, Li~Wan, Philip~Andrew Mansfield, and Ignacio~Lopz
  Moreno,
\newblock ``{Speaker diarization with LSTM},''
\newblock in {\em IEEE ICASSP}, 2018, pp. 5239--5243.

\bibitem{likas2003global}
Aristidis Likas, Nikos Vlassis, and Jakob~J Verbeek,
\newblock ``The global k-means clustering algorithm,''
\newblock {\em Pattern recognition}, vol. 36, no. 2, pp. 451--461, 2003.

\bibitem{PIC}
Wei Zhang, Deli Zhao, and Xiaogang Wang,
\newblock ``Agglomerative clustering via maximum incremental path integral,''
\newblock {\em Pattern Recognition}, vol. 46, no. 11, pp. 3056--3065, 2013.

\bibitem{SSPIC}
Prachi Singh and Sriram Ganapathy,
\newblock ``Self-supervised representation learning with path integral
  clustering for speaker diarization,''
\newblock {\em IEEE/ACM Transactions on Audio, Speech, and Language
  Processing}, vol. 29, pp. 1639–1649, 2021.

\bibitem{shinji2019ASRU}
Yusuke Fujita et~al.,
\newblock ``{End-to-End Neural Speaker Diarization with Self-attention},''
\newblock in {\em ASRU}, 2019, pp. 296--303.

\bibitem{shinji2020Interspeech}
Shota Horiguchi, Yusuke Fujita, Shinji Watanabe, Yawen Xue, and Kenji
  Nagamatsu,
\newblock ``{End-to-End Speaker Diarization for an Unknown Number of Speakers
  with Encoder-Decoder Based Attractors},''
\newblock in {\em {INTERSPEECH}}, 2020.

\bibitem{horiguchi2021towards}
Shota Horiguchi, Shinji Watanabe, Paola Garc{\'\i}a, Yawen Xue, Yuki Takashima,
  and Yohei Kawaguchi,
\newblock ``Towards neural diarization for unlimited numbers of speakers using
  global and local attractors,''
\newblock in {\em IEEE ASRU}, 2021, pp. 98--105.

\bibitem{Lin2019}
Qingjian Lin, Ruiqing Yin, Ming Li, Hervé Bredin, and Claude Barras,
\newblock ``{LSTM Based Similarity Measurement with Spectral Clustering for
  Speaker Diarization},''
\newblock in {\em Proc. {INTERSPEECH}}, 2019, pp. 366--370.

\bibitem{singh2021self}
Prachi Singh and Sriram Ganapathy,
\newblock ``Self-supervised metric learning with graph clustering for speaker
  diarization,''
\newblock in {\em IEEE ASRU}, 2021, pp. 90--97.

\bibitem{HGNN}
Yifan Xing, Tong He, Tianjun Xiao, Yongxin Wang, Yuanjun Xiong, Wei Xia, David
  Wipf, Zheng Zhang, and Stefano Soatto,
\newblock ``Learning hierarchical graph neural networks for image clustering,''
\newblock in {\em Proc. IEEE ICCV}, 2021, pp. 3467--3477.

\bibitem{hamilton2017inductive}
Will Hamilton, Zhitao Ying, and Jure Leskovec,
\newblock ``Inductive representation learning on large graphs,''
\newblock {\em Advances in neural information processing systems}, vol. 30,
  2017.

\bibitem{GNNfordiarization}
Jixuan Wang et~al.,
\newblock ``Speaker diarization with session-level speaker embedding refinement
  using graph neural networks,''
\newblock in {\em IEEE ICASSP}, 2020, pp. 7109--7113.

\bibitem{kipf2016semi}
Thomas~N Kipf and Max Welling,
\newblock ``Semi-supervised classification with graph convolutional networks,''
\newblock {\em arXiv preprint arXiv:1609.02907}, 2016.

\bibitem{TongGCN}
Fuchuan Tong et~al.,
\newblock ``Graph convolutional network based semi-supervised learning on
  multi-speaker meeting data,''
\newblock in {\em IEEE ICASSP}, 2022, pp. 6622--6626.

\bibitem{zeinali2019but}
Hossein Zeinali et~al.,
\newblock ``{BUT} system description to voxceleb speaker recognition challenge
  2019,''
\newblock {\em arXiv preprint arXiv:1910.12592}, 2019.

\bibitem{mccowan2005ami}
I~McCowan, J~Carletta, W~Kraaij, S~Ashby, S~Bourban, M~Flynn, M~Guillemot,
  T~Hain, J~Kadlec, V~Karaiskos, et~al.,
\newblock ``The {AMI} meeting corpus,''
\newblock in {\em International Conference on Methods and Techniques in
  Behavioral Research}, 2005, pp. 137--140.

\bibitem{nagrani2017voxceleb}
Arsha Nagrani, Joon~Son Chung, and Andrew Zisserman,
\newblock ``Voxceleb: A large-scale speaker identification dataset,''
\newblock {\em Proc. of {INTERSPEECH}}, pp. 2616--2620, 2017.

\bibitem{Chung2018}
Joon~Son Chung, Arsha Nagrani, and Andrew Zisserman,
\newblock ``Voxceleb2: Deep speaker recognition,''
\newblock in {\em Proc. of {INTERSPEECH}}, 2018, pp. 1086--1090.

\bibitem{panayotov2015librispeech}
Vassil Panayotov et~al.,
\newblock ``Librispeech: an asr corpus based on public domain audio books,''
\newblock in {\em IEEE ICASSP}, 2015, pp. 5206--5210.

\bibitem{chung20_interspeech}
Joon~Son Chung et~al.,
\newblock ``{Spot the Conversation: Speaker Diarisation in the Wild},''
\newblock in {\em Proc. INTERSPEECH 2020}, 2020, pp. 299--303.

\bibitem{ryant21_interspeech}
Neville Ryant et~al.,
\newblock ``{The Third DIHARD Diarization Challenge},''
\newblock in {\em Proc. {INTERSPEECH}}, 2021, pp. 3570--3574.

\bibitem{landini2020but}
Federico Landini et~al.,
\newblock ``{BUT} system for the second {DIHARD} speech diarization
  challenge,''
\newblock in {\em IEEE ICASSP}, 2020, pp. 6529--6533.

\bibitem{spectral_clustering}
Andrew Ng, Michael Jordan, and Yair Weiss,
\newblock ``On spectral clustering: Analysis and an algorithm,''
\newblock {\em Advances in neural information processing systems}, vol. 14,
  2001.

\bibitem{landini2020bayesian}
Federico Landini et~al.,
\newblock ``Bayesian {HMM} clustering of x-vector sequences ({VBx}) in speaker
  diarization: theory, implementation and analysis on standard tasks,''
\newblock {\em arXiv preprint arXiv:2012.14952}, 2020.

\bibitem{dawalatabad2021ecapa}
Nauman Dawalatabad et~al.,
\newblock ``{ECAPA-TDNN} embeddings for speaker diarization,''
\newblock {\em arXiv preprint arXiv:2104.01466}, 2021.

\bibitem{landini2021analysis}
Federico Landini et~al.,
\newblock ``Analysis of the {BUT} diarization system for voxconverse
  challenge,''
\newblock in {\em IEEE ICASSP}, 2021, pp. 5819--5823.

\bibitem{wang2022similarity}
Weiqing Wang et~al.,
\newblock ``Similarity measurement of segment-level speaker embeddings in
  speaker diarization,''
\newblock {\em IEEE/ACM Transactions on Audio, Speech, and Language
  Processing}, vol. 30, pp. 2645--2658, 2022.

\bibitem{anguerabeamformit}
X.~Anguera, C.~Wooters, and J.~Hernando,
\newblock ``Acoustic beamforming for speaker diarization of meetings,''
\newblock {\em IEEE Transactions on Audio, Speech, and Language Processing},
  vol. 15, no. 7, pp. 2011--2021, 2007.

\end{thebibliography}
\end{document}